\documentstyle[preprint,aps]{revtex}

\begin{document}
\draft
\title{{\large {\bf The stochastic gravitational background from 
inflationary phase transitions}}}
\author{Carlo Baccigalupi$^{1,2}$, Luca Amendola$^{2}$, 
Pierluigi Fortini$^{1}$ \& Franco Occhionero$^{2}$}

\address{$^{1}$ INFN and Dipartimento di Fisica, 
\\ Universit\`a di Ferrara, Via del Paradiso 12, 44100 Ferrara, Italy\\
$^2$Osservatorio Astronomico di Roma,
Viale del Parco Mellini 84, 00136 Roma, Italy}
\maketitle

\maketitle
 
\baselineskip 10pt
\vspace{1.cm} 
\begin{abstract}
We consider true vacuum bubbles generated in a first order phase
transition occurring during the slow rolling era of a two field
inflation: it is known that gravitational waves are produced by the
collision of such bubbles.
We find that the epoch of the phase transition strongly
affects the characteristic peak frequency of the gravitational waves,
causing an observationally interesting redshift in addition to the
post-inflationary expansion. In particular it is found that
a phase transition occurring
typically 10$\div$20 $e-$foldings before the reheating
at $kT\simeq 10^{15}$ GeV
may be detected by the next Ligo gravity waves
interferometers.

Moreover, for recently proposed models capable of generating
the observed large scale voids as remnants of the primordial bubbles 
(for which the characteristic wave lengths are several tens of Mpc), 
it is found that the level of 
anisotropy of the cosmic microwave background
provides a deep insight upon the physical
parameters of the effective Lagrangian.

\end{abstract}

	\section{\normalsize\bf Introduction}
	
The knowledge of the nature of the matter perturbations 
in the observed universe is crucial for obtaining information
about the very early universe and the very high energy physics.
The deepest presently available window to the early universe
is the cosmic microwave background (CMB), that in the next decade
will be deeply investigated by the Map and Planck missions,
after the discovery of its anisotropies by COBE \cite{COBE}.

However, the cosmic gravitational background (CGB) will play
a crucial role in the next future as the most powerful tool in
reconstructing primordial physics (see \cite{A} for an extensive
overview). Infact, 
it consists of the gravitational waves (GW) generated during the 
inflationary era; they are carriers 
of unperturbed physical traces of the very primordial
history of the universe, since the GW decoupling probably 
occurred about 70 $e-$foldings before the recombination! 

The study of the perturbations and defects produced during
inflation underwent recently a great revival; it is motivated by the
evidence of strong inhomogeneities from the direct reconstruction of 
the three dimensional matter distribution 
traced by galaxies and their peculiar velocities in the modern 
redshift surveys. The most important conclusions of these observations 
are that strong underdensities, or voids, appear very prominent
in the data \cite{AV}\cite{EPD1}\cite{EPD2}\cite{DFWGHS}. 
In search for the inflationary generation mechanisms for 
such inhomogeneities, one of the most interesting ideas introduced 
in cosmology in recent years is the possibility 
of performing a phase transition {\it during} inflation.
In such scenarios, two fields act on stage: one, say $\omega$, slow
rolls, driving enough inflation to solve the standard problems;
the second field, say $\psi$, tunnels from a false vacuum state to an
energetically favoured true vacuum state, producing
bubbles of the new phase embedded in the old one. 
Both processes are governed by a two-field
potential $U(\omega ,\psi)$.
To avoid the graceful exit
problem, the true vacuum state has to allow for a period of inflation
on its own.
We can then speak of a true vacuum channel over which the
bubbles slow roll until inflation ends, and reheating takes place.
Depending on the potential, we can distinguish two different 
scenarios of first order inflation.
The first is the classical extended inflation
\cite{LS,LSB,AF}: the bubbles
are produced in a copious quantity, so that they eventually
fill the space and complete the transition. To avoid
too large distortions on the CMB, this scenario must produce 
very small bubbles \cite{TWW,LW},
so that they are rapidly thermalized after inflation.
No trace of the bubbles is left in our universe, and from this
point of view such scenarios do not lead to new predictions over
inflation without bubble production.
In the second scenario, proposed in \cite{OA} and implemented in
\cite{ABKOR}, the phase transition is completed {\it before} 
the end of inflation; the amount of $e$-foldings between the
phase transition and the end of inflation makes the scale of the 
bubbly perturbations interestingly non-vanishing, and allows
them to leave observable traces. In particular, if a
phase transition occurred sufficientely early,
the bubbles are streched to cosmological scales, and
the present large scale structure is therefore strictly
linked to the primordial originating transition,
which is therefore observable and testable
as firstly suggested in \cite{LA}.

In this work we concentrate upon the traces of the above
phaenomenology on the CGB.
The most natural way for the nucleated bubbles to generate
gravitational waves is through collision with each other;
compared with the ordinary tensor perturbations occurring 
in slow rolling inflation, this is infact a very potent 
source of primordial GW, as firstly argued in \cite{TW}.
The problem was analysed in the context of the ordinary
extended inflation, in which the bubbles are completely empty
and the phase transition occurs at the end of inflation.
The computation of the amplitude and frequency spectrum
of the gravitational radiation was performed firstly
in the case of two bubbles \cite{KTW} and then by considering
an envelope of hundreds of bubbles \cite{KT}, that substantially
confirmed the previous results; a work emphasising the observation 
possibilities can be found in \cite{KTW2}. Finally in \cite{KKT} the
problem of gravitational waves from lower energy first order 
phase transition (like the electro-weak transition occurring nearly 
at 100 GeV) was considered.

Here we want to extend these results to the second kind of 
inflationary phase transitions mentioned above; the 
main differences from the treated cases are that $i$) 
the nucleation epoch occurs before the end of inflation, and $ii$)
generally the bubble are not empty.

The paper is organized as follows: in Section II we recall the
main results on the gravitational radiation emitted during
the first order phase transition in the scenario of extended 
inflation, pointing out the approximations for which the computations 
and the results are valid; 
in Section III we extend these results to recently proposed 
inflationary models 
capable of performing a first order phase transition before
the end of inflation and we discuss the
observation possibilities and the existing constraints; 
finally, Section IV contains the conclusions.

\section{Collisions at the end of inflation}

A detailed analysis of a first order phase transition
in the context of extended inflation
can be found in \cite{TWW}.
The field $\psi$ that undergoes the transition is the
same that drives inflation, and its dynamics is assumed to
be governed by a potential containing two non-degenerate minima,
the true and false vacuum (TV,FV).
The central quantity needed to characterize the transition 
is the bubble nucleation rate for unit volume
in the semiclassical limit \cite{C}
\begin{equation}\label{tunnrate}
\Gamma={\cal M}^4 e^{-B}\ \ ,
\end{equation}
where $B$ is the Euclidean least action over the bounce minus
the action for the external deSitter spacetime solution \cite{C}.
The constant ${\cal M}$ (with a dimension of mass)
is of the order of the energy $T$ at which the phase transition occurs.
Values of $\cal M$ a few orders of magnitude below the Planck 
mass $m_{PL}$ are generally assumed, in order to avoid quantum 
gravity effects \cite{GZ}.

The transition is completed roughly when at least one bubble per
unit Hubble volume is nucleated, and accurate computations \cite{TWW}
show that extended inflation is successful (in the sense that
bubbles percolate) if at a time $t_{e}$
\begin{equation}\label{condition}
Q={4\pi\over 9}\left({\Gamma\over H^{4}}\right)_{t_{e}}=1\ \ .
\end{equation}
The bubbles are empty and their walls rapidly approach the light speed.
Depending on the intensity of the nucleation rate, the collisions
between the bubbles occur when they have comoving spatial dimension
less or equal to the effective horizon $H^{-1}$ at the transition
epoch given by $T\simeq\cal M$. In the context of the standard
extended inflation almost all the bubbles are nucleated
at the end of inflation, so that if we take
$H_{0}=100h$Km/sec/Mpc in an $\Omega=1$ universe their
comoving size is approximatively $10^{-21}{\ \rm {\it h}^{-1} Mpc}$. 
Such bubbles soon reenter the horizon and rapidly thermalize without 
leaving trace in the matter distribution.

On the other hand, a very characteristic GW spectrum is
produced during the transition.
If the bubbles are perfectely spherical, no spacetime perturbation
is seen from the outside. When two (or more) bubbles collide
they become a GW source, and the collection of collisions during 
the whole phase transition leaves observationally interesting
traces on the CGB \cite{TW}. 
The problem of computing the emitted GW spectrum
is conceptually simple but computatively difficult. 
One has to simulate the bubble nucleations with rate given by
(\ref{tunnrate}) and to evolve them with the Klein-Gordon equation.
Whenever collisions occur, one must compute the radiation emitted.
It has been performed \cite{KTW} with the aid of two main 
approximations: linearized gravity and static background.
The former is consistent if the fraction of energy that goes
in GW is small with respect to unity; the latter is valid if the 
transition completes within an Hubble time.
Following \cite{W}, the total energy radiated at frequency $\omega$ 
in the direction ${\bf k}$ into the solid angle $d\Omega$ 
and in the interval $d\omega$ is
\begin{equation}\label{GW}
{dE\over d\omega d\Omega}=2G\omega^{2}\Lambda_{ijlm}({\bf k})
T^{ij*}({\bf k},\omega)T^{lm}({\bf k},\omega)\ \ ,
\end{equation}
where $\Lambda_{ijlm}({\bf k})$ is the projection tensor for
gravitational radiation \cite{W} and $T^{\mu\nu}({\bf k},\omega )$ 
the Fourier transform of the energy momentum tensor.
The computations were firstly performed in the simplified case
of two colliding bubbles \cite{KTW}. 
The end of the transition in that case was
modelled as a modulating function $T_{ij}(t)\rightarrow T_{ij}(t)C(t)$
that makes the signal vanish after a cutoff time $\tau$ of the order 
of the initial separation between the bubbles. In \cite{KT}
an ingenious model for $C(t)$ was introduced by excluding from
the integration the spatial region where bubbles
overlap; at $t\gg t_{e}$ the bubbles completely fill the space,
and the GW emission ends. This also allowed to perform many bubbles
simulations. The results are intuitive and consistent
with the approximations. Firstly, the bounce is 
approximated as \cite{TWW}
\begin{equation}\label{beta}
B(t)\simeq B(t_{e})-\beta (t-t_{e})\ \ ,
\end{equation}
where $\beta^{-1}$ sets the natural scale for the phase transition.
Since this is of the order of the initial average separation between
the colliding bubbles, the static backgorund approximation requires
that $\beta^{-1}\le H^{-1}$. 
The GW spectrum shows a very characteristical peak at
$\omega_{GW}\simeq 1.6\beta$
and the fraction of energy radiated in GW is found
$\Omega_{GW}=\rho_{GW}/ \rho_{c}\simeq 0.06\left(H/\beta\right)^{2}$;
$\rho_{c}=3H^{2}/8\pi G=g\pi^{2}T_{e}^{4}/30$ 
is the critical energy density
at the time $t_{e}$; $g$ counts the number of relativistic
degrees of freedom at the temperature $T_{e}$ of the phase transition,
and is typically of the order of 100. We remark that these results 
agree with physical expectations: the peak frequancy is simply the 
inverse of the time scale of the process, and the $(H/\beta )^{2}$ 
scaling for $\Omega_{GW}$ can be naturally inferred in terms of the 
energy density $\rho_{c}$ and $\beta^{-1}$ \cite{KTW}. To obtain the 
corresponding present quantities, one has to take into account the 
cosmic redshift from $T_{e}$ down to the present 3 K. $\Omega_{GW}$
remains unchanged during the radiation epoch, but
undergoes a decrease during the matter dominated era;
$\omega_{GW}$ redshifts via the cosmic expansion. 
The quantities relevant at the present are therefore
\begin{equation}\label{omeganow}
\omega_{GW}({\rm Hz})\simeq 3\cdot 10^{7}\left({\beta\over H}\right)
g_{100}^{1/6}T_{15}\ \ ,
\end{equation}
\begin{equation}\label{omegaGWnow}
\Omega_{GW}h^{2}\simeq 10^{-6}\left({H\over\beta}\right)^{2}
g_{100}^{-1/3}\ \ ,
\end{equation}
where we have defined $T_{15}=T_{e}/(10^{15} {\rm GeV})$ and 
$g_{100}=g/100$.

We emphasize once again that the above results hold for the extended
inflation scenario, in which bubbles are nucleated at the end of
inflation and rapidly thermalize.
In the next section we will extend these results to the models
of first order inflation capable to perform the nucleation epoch 
well before the end of inflation,
showing how the above results change and discussing the consequences
on the observability.

\section{Collisions during inflation}

Recent studies on inflation \cite{OA} have shown that a not $ad\ hoc$
slow rolling is generated by the gravitational part of
the Lagrangian, already at the level of Fourth Order Gravity (FOG).
A matter field potential containing false and true vacuum minima
generates a first order phase transition, but the latter 
in general occurs at an epoch characterized by a number $N_{0}$ 
of $e-$foldings before the end of inflation. 
Depending on this quantity, the scale of the 
bubble-like perturbations may be stretched out to cosmological size 
by the exponential growth, thus candidating the bubbles to be the seeds 
for the formation of the voids observed today \cite{LA,AO,ABKOR}. 
We consider the consequences on the CGB spectrum of this FOG model
(and we refer to \cite{OA} and \cite{ABKOR} for an extensive 
treatment); since a conformal transformation makes the Lagrangian as
in ordinary two field inflation, our results hold for any kind of
inflation in which one field slow rolls and the other one undergoes a 
first order phase transition.

Once the conformal transformation is performed, the FOG action takes  
the form
\begin{equation}\label{eua}
S=\int\sqrt{-g} d^4x \left[ -{{\cal R}\over 16\pi}
+{3\over 4\pi}\omega_{;\mu}\omega^{;\mu}+
{1\over 2} e^{-2\omega}\psi_{;\mu}\psi^{;\mu}
+U(\psi,\omega)\right]\,,
\end{equation}
where the potential is
\begin{equation}\label{conpot}
U(\psi,\omega)=e^{-4\omega}\left[V(\psi)+{3M^2\over 32\pi}
W(\psi)\left( 1-e^{2\omega}\right)^2  \right]\,,
\end{equation} 	
and generates TV bubbles with
$W(\psi)=1+(8\lambda/\psi_0^4)\psi^2
(\psi-\psi_0)^2$, a degenerate quartic, and
$V(\psi)=(1/2)m^{2}\psi^{2}$, the symmetry breaking term.
The slow roll inflation driven by $\omega$ takes place at 
$\omega\gg 1$, and is over when $\omega$ approaches zero.
At large $\omega$, the potential $U$ is dominated by $W(\psi)$,
and thus the false vacuum minimum at 
$\psi\approx \psi_0$, for which $U_{FV}=e^{-4\omega}[(3M^2/32\pi)
(1 - e^{2\omega})^2+ V (\psi_0)]$
is unstable with respect to tunneling towards
the true vacuum
$\psi=0$ (for which $U_{TV}\approx
3M^2/32\pi$). At small $\omega$, instead, $U$ is dominated by
$V(\psi)$, and both the true and the false vacua converge
to the global zero-energy minimum at $\omega =\psi=0$,
where inflation ends and  reheating takes place.
The slow-roll solution in this model for $\omega\gg 1$ 
can be written very conveniently as \cite{OA} $N =(3/4)e^{2\omega}$
where $N$ is the number of $e$-foldings to the end of inflation.

It has been shown \cite{ABKOR} that taking into account the 
gravitational corrections and dropping
the thin wall approximation due to a finite thickness $\delta R$ 
for the wall surrounding a bubble of radius $R$, the nucleation 
rate $\Gamma$ defined in (\ref{tunnrate}) becomes
\begin{equation}\label{tunnrate2}
\Gamma ={{\cal M}^{4}B^{2}(\delta R)^{4}\over 16 R^{4}}e^{-B}=
{\cal M'}^{4}e^{-B}\ \ ,
\end{equation}
and in terms of $N$ the bounce is
\begin{equation}\label{bn}
B={N^4\over N_1^4}\left[1-\left({N_2\over N}
\right)^2\right]\left[1-\left({N\over N_3}
\right)^4\right]
\end{equation}
where the first brackets is the thin wall correction, 
and the second brackets is the gravitational correction;
moreover, the $N_{i}$ are related to the physical quantities by
\begin{equation}\label{n}
N_1^2 = {3^{3/2}\over 4}{\psi_0 m^3\over M^2\lambda}\ ,\ 
N_2^2 = {27\pi\over 8}{\psi_0^2 m^2\over M^2\lambda}\ ,\
N_3^2 = \left({27\pi \over 32}\right)^{1/2} 
{\psi_0 m^2\over M^2\lambda^{1/2}}\nonumber\ \ .
\end{equation}

By indicating with $N_{0}$ the amount of $e-$foldings to the end
of inflation when the phase transition takes place, the quantity $Q$
defined in (\ref{condition}) takes the approximate form  
\begin{equation}\label{q}
Q={4\pi \Gamma \over 9H^4}=\exp \left\{
{(N_0^4-N^4)\over N_1^4}\left[1-\left({N_2\over N_0}
\right)^2\right]\left[1-\left({N_0\over N_3}
\right)^4\right]\right\}\ \ .
\end{equation}
The two approximations that have been adopted to obtain the above 
formulas are satisfied by the conditions $N>N_2$ (thin-wall), 
and for $N<N_3$ (gravitational correction). Also, $N_{0}>N_{1}$
is required to guarantee tunneling through TV bubbles \cite{OA}.

The central underdensity of a bubble is determined by the shape of 
the potential (\ref{conpot}) at the nucleation epoch, characterizes 
by $N$:
\begin{equation}\label{delta}
\delta\equiv |\delta\rho/\rho|=
{U_{FV}-U_{TV}\over U_{FV}}=[(N/N_4)^2+1]^{-1}\ \ ,\ \  
N_4^2=3\pi {\psi_0^2 m^2\over M^2}={64\over 9\pi}
M^{2}N_{1}^{-4}N_{3}^{8}\ \ .
\end{equation}

From the current microwave
background measurements, we obtain $M\approx 5\cdot 10^{-6}$
(in Planck units) \cite{OA}, a value that we will adopt in the 
results below. However,
since in our model one should also consider
the contribution of the bubbles to the microwave background, this
constraint is actually only an upper limit on $M$.
As an intuitive feature, note that the four physical parameters
$M,m,\lambda ,\psi_{0}$, and therefore the four $e$-folding constants 
$N_{1-4}$ fully determine the inflationary potential at $N_{0}$ 
(the nucleation epoch):
the vacuum energy density, the energy difference between FV and TV,
the amplitude of the barrier and the value $\psi_{0}$ of the FV phase. 

The linearized gravity approximation involved to obtain the results
(\ref{omeganow},\ref{omegaGWnow}) is surely satisfied in our case
since the energy that goes into the walls and therefore in GW is a 
fraction $\delta$ given by (\ref{delta}) of the total FV energy 
in the bubble. The other important approximation is to consider
spacetime as static during the transition; in other words, the time 
scale of the nucleation era must be smaller than the Hubble time. 
This is satisfied in our case if the condition $\beta^{-1}\le H^{-1}$ 
holds at $N_{0}$, where $\beta$ is defined in (\ref{beta});
this means $-(\partial B/\partial t)_{N_{0}}\ge H$, that using 
(\ref{bn}) becomes
\begin{equation}\label{fasttrans}
{\cal B}=\left({\partial B\over\partial N}\right)_{N_{0}}=
{4N_{0}^{3}\over N_{1}^{4}}
\left[1-{1\over 2}\left({N_{2}\over N_{0}}\right)^{2}-
2\left({N_{0}\over N_{3}}\right)^{4}+
{3\over 2}{N_{0}^{2}N_{2}^{2}\over N_{3}^{4}}\right]\ge 1\ \ .
\end{equation}
By defining 
\begin{equation}\label{xy}
x=(N_{2}/N_{0})^{2}\ \ ,\ \ y=(N_{0}/N_{3})^{4}\ \ ,
\end{equation} 
the quantity in brackets in (\ref{fasttrans}) is simply
\begin{equation}\label{z} 
z(x,y)=1-{x\over 2}-2y+{3xy\over 2}\ \ ,
\end{equation}
still taking into account the gravitational and post-thin wall 
corrections. The range of interest for both $x$ and $y$ 
is $[0,1]$ for consistency of our approximations. Since $N_{1}$ does
not appeare in $z(x,y)$, for the validity of (\ref{fasttrans}) it
is enough that $z(x,y)$ assumes some positive value; it is easily
seen that this is true in the range $0<x<1,\ \ 0<y<(x-2)/(3x-4)$. 
By choosing a value of $N_{0}$ and $N_{2}<N_{0}$ to fix $x$, a value
of $y$ always exists that yields $z(x,y)>0$; therefore, our model 
performs a fast transition for
\begin{equation}\label{fasttrans2}
N_{1}\le [4N_{0}^{3} z(x,y)]^{1/4}\ \ .
\end{equation}

We can now extend the results (\ref{omeganow},\ref{omegaGWnow})
to the present inflationary models. The peak frequency of the
GW spectrum contains an $additional$ redshift due to the inflationary
expansion between the phase transition and the end of slow rolling. 
Moreover, in calculating $\Omega_{GW}$ we must take into account that 
the fraction of energy that goes in gravitational radiation is reduced 
by a fraction $\delta$, (\ref{delta}), with respect to 
(\ref{omegaGWnow}). Therefore, the results are
\begin{equation}\label{gwfog1}
\omega_{GW}({\rm Hz})\simeq 3\cdot 10^{7}{\cal B}
g_{100}^{1/6}T_{15}\exp({-N_{0}})\ \ ,
\end{equation}
\begin{equation}\label{gwfog2}
\Omega_{GW}h^{2}\simeq 10^{-6}{\cal B}^{-2}
g_{100}^{-1/3}\delta (N_{0})\ \ .
\end{equation}
In the following we will adopt the values $T_{15}\simeq 1$ 
(fixing the epoch of the end of slow rolling at typical GUT 
energy scales), and
$g_{100}\simeq 1$.

From the above formulas it is clear that in general
any model of two field first order inflation radically changes
the frequencies of the GW spectrum to be observed
today with respect to the ordinary extended inflation models. 
The main difference is represented by the shift of the peak 
frequency towards lower values depending on the amount of
$e-$foldings between the phase transition and the end of inflation.
This has very interesting consequences for what concerns the search
for the present traces of the era driven by very high energy physics. 
We make now some general considerations about this,
referring to the next two sub-sections to a detailed analysis of
two important cases.
 
Formulas (\ref{gwfog1},\ref{gwfog2}) may be rewritten emphasizing 
the relation between the physical parameters
\begin{equation}\label{gwfog3}
\Omega_{GW}h^{2} = 9\cdot 10^{8}
{\delta (N_{0})\over[\omega_{GW}({\rm Hz})]^{2}}
\exp (-2N_{0})\ \ ,
\end{equation}
with the constraint 
${\cal B}=10^{-3}\sqrt{\delta (N_{0})/\Omega_{GW}h^{2}}\ge 1$.
In Fig.1 we plot experimentally interesting 
$\Omega_{GW}h^{2}$ as a function of the peak frequency 
$\omega_{GW}$ and for different $N_{0}$ 
(the change in grey tonality corresponds to a scansion of 5 in 
$N_{0}$, as indicated on the legend box); 
for clearness, we have set $\delta_{N_{0}}=.1$ (top) 
and .01 (bottom). Each point on the figure correspond to
the GW spectrum generated by a first order phase transition 
occurred $N_{0}$ $e$-foldings before the end of inflation.
For $N_{0}\rightarrow 0$ we obtain the known results
in the case of extended inflation 
(\ref{omeganow},\ref{omegaGWnow}).
The increase of $N_{0}$ pushes the peak frequency
towards exponentially small values and 
intersects very interestingly the plotted expected 
level of sensitivities of the next generation 
of GW interferometric detectors: if first order
phase transitions occurred at $10\le N_{0}\le 30$
during the inflationary era 
(realistically arising from spontaneous breaking of very 
high energy symmetries), they should leave next future
detectable peaks in the CGB frequency spectrum. 
By continuing to increase $N_{0}$ beyond several 
decades, we enter in the class of phase transitions that could leave 
traces in the present large scale matter distribution. In particular, 
at $N_{0}\ge 50$, the remnants of the nucleated bubbles are
of astrophysically interesting size, and correspond to the large 
voids detected in the galaxy distribution \cite{LA,OA,AO,OBAM}
\cite{EPD1,EPD2}. 
For such models, the GW frequencies correspond to cosmological 
wavelenghts, to be interestingly investigated through their induced 
CMB anisotropies; as we will show in the sequel, a deep insight 
already comes from considering the anisotropy amplitude; however it is 
interesting to point out that in the future these models could 
be tested with powerful methods (presently under investigation) 
capable to extract the pure CGB signal from the data coming from 
the next high resolution CMB experiments \cite{KK,SZ}.

In the following we consider in more detail the correspondence
between the CGB and the parameters of the effective Lagrangian,
focusing on two important cases: 
spectra detectable by the next Ligo 
interferometric detectors, and spectra 
from colliding bubbles
of astrophysically interesting size.

\subsection{Next future detectable GW from bubbles}
 
The Advanced Ligo project will reach a sensibility of about 
$\Omega_{GW}h^{2}\simeq 10^{-11}$ in the frequency range 
$\omega_{min}\simeq 1{\ \rm Hz},\ \omega_{max}\simeq 10^{2}$ Hz
(see \cite{A} and references therein). 
From (\ref{gwfog1},\ref{gwfog2}), 
we see that the amplitude and the peak frequency fix the products
${\cal B}\exp (-N_{0})$ and ${\cal B}^{-2}\delta (N_{0})$, namely
two conditions for our physical parameters. Then, by fixing 
$N_{2},\ N_{3}$ so to yield a reasonable parametric region
for $N_{0},\ N_{1}$ (we remember that $N_{2}\le N_{0}\le N_{3}$),
we may search for the physical parameter set that could explain the
observation of a definite peak at some frequency in the above 
ranges. Such set is constrained by (\ref{gwfog1}),
\begin{equation}\label{l0}
\left[{3\cdot 10^{7}\over \omega_{max}}4N_{0}^{3}
z(x,y)e^{-N_{0}}\right]^{1\over 4}\le N_{1}\le
\left[{3\cdot 10^{7}\over \omega_{min}}4N_{0}^{3}
z(x,y)e^{-N_{0}}\right]^{1\over 4}\ \ ,
\end{equation}
to have the peak frequency in the observation range, 
and by (\ref{gwfog2}),
\begin{equation}\label{l1}
N_{1}\ge\left[{9\pi z^{2}(x,y)N_{0}^{8}\over 
8M^{2}N_{3}^{8}}\Omega_{GW}h^{2}+
\sqrt{\left({9\pi z^{2}(x,y)N_{0}^{8}\over 
8M^{2}N_{3}^{8}}\Omega_{GW}h^{2}\right)^{2}+
16z^{2}(x,y)N_{0}^{6}\, 10^{6}\Omega_{GW}h^{2}}\right]^{1\over 4}\ \ ,
\end{equation}
to have a sufficient amplitude. 
Moreover (\ref{fasttrans2}) (with $z(x,y)>0$ of course)  
must be satisfied to have the completion of the transition 
in less than an Hubble time and $N_{1}<N_{0}$ to have 
nucleation of TV bubbles. 
The dashed areas in Fig.2 are the regions of the 
parameters $(N_{0},N_{1})$ for which both the peak frequency and 
the amplitude fall into the observation ranges; the upper panel shows 
the limit in which the post-thin wall and gravitational corrections 
are negligible, while in the lower panel such corrections are 
important. The constraints define the region of observability shaded 
in Fig.2; as an interesting and intuitive feature 
we note that a first order inflationary phase transition, 
that is Ligo detectable,
occurs at $N_{0}=10\div 20$ $e-$foldings before the end of inflation, 
that is just the amount necessary to redshift the corrisponding 
frequencies in ordinary extended inflation ($\simeq 10^{7}\ {\rm Hz}$, 
see (\ref{omeganow}) and Fig.1) towards the observation 
range. From to the nucleation epoch it is easily found that the 
comoving size of these perturbations is approximatively 
$10^{-18}\div 10^{-14}{\ \rm {\it h}^{-1} Mpc}$; this means that 
they reenter the horizon early in the radiation 
dominated era and most likely do not leave traces in the density field. 
Finally we remark once again that for each set of $N_{1-4}$ 
there exists a set of physical parameters 
${\cal M'},\psi_{0},m,\lambda$, 
as it is easily seen from (\ref{tunnrate2},\ref{bn}, \ref{n},\ref{q}).

\subsection{Large scale bubbles}

As we already mentioned in the Introduction, the recent analysis on 
the modern redshift surveys have assessed the large scale voids as
the dominant feature of the nearby universe. Several works
have been performed in these years on the hypothesis that the
primordial origin of the voids is in an inflationary phase transition
in the context of models of FOG and in general two field inflation
\cite{OA,ABKOR,LA,AO,OBAM,BAO,BAO2}. Here we add a new tool.
It has been shown \cite{OA} that voids of tens of Mpc of diameter
would be the relics of bubbles with comoving size expanded by a
factor $\exp (50)$ between the phase transition and the end of 
inflation. This request fixes the product 
\begin{equation}\label{g}
{\cal B}^{-1}\exp (N_{0})\simeq \exp (50)\ \ ,
\end{equation} 
and by requiring ${\cal B}\ge 1$ we have $N_{0}\ge 50$.
The frequency in (\ref{gwfog1}) is therefore:
\begin{equation}\label{void1}
\omega_{GW} ({\rm Hz})\simeq 10^{-15}\ \ .
\end{equation} 
On these frequencies there is the most stringent presently 
available constraint upon the amplitude of the CGB,  
provided by the amount of anisotropy of the CMB (see \cite{A} and
references therein). Precisely, this constraint imposes that
\begin{equation}\label{allen}
\Omega_{GW}h^{2}\le 7\cdot 10^{-11}\left({H_{0}\over\omega}\right)^{2}
\ \ {\rm for}\ \ H_{0}\le \omega \le 30H_{0}\ \ ,
\end{equation}
and consequentely our frequencies localize themselves on the high
frequency border of this range, where $\Omega_{GW}h^{2}\le 10^{-13}$
($H_{0}=100h$ Km/sec/Mpc). Note that the corrisponding wavelenghts 
determine CMB anisotropies on small angular scales 
($\theta < 1^{o}$), that will be deeply investigated by the Map and
Planck missions of the next decade, as mentioned in the Introduction. 
A great effort is being done in order to develope theoretical 
analysis instruments capable to extract the pure CGB signal from the
whole spectrum of CMB anisotropies, both for what concerns pure 
$\delta T/T$ and polarization (see e.g. \cite{SZ,KK}); 
this matter will provide a powerful tool to 
investigate the whole CGB spectrum,
and in particular the traces of early inflationary phase 
transitions, that are the subject here. 
Here we concentrate on the overall
amplitude constraint given by (\ref{allen}), since it already 
provides a deep insight into the physical parameter space; 
from (\ref{delta},\ref{gwfog2}) we see that it must be 
\begin{equation}\label{void2}
\delta (N_{0})\le 10^{-7}{\cal B}^{2}\ \ ,
\end{equation}
or, alternatively to (\ref{l0}),
\begin{equation}\label{void3}
N_{1}\le\left[{9\pi z^{2}(x,y)N_{0}^{8}\over 
8\cdot 10^{7}M^{2}N_{3}^{8}}+
\sqrt{\left({9\pi z^{2}(x,y)N_{0}^{8}\over 
8\cdot 10^{7}M^{2}N_{3}^{8}}\right)^{2}+
{16z^{2}(x,y)N_{0}^{6}\over 10^{7}}}\right]^{1\over 4}\ \ ;
\end{equation}
this sets a very narrow range (at fixed $N_{2}$ and $N_{3}$) 
for $N_{0}$ and $N_{1}$. Moreover, another condition
arises from (\ref{delta}) for the minimal $\delta (N_{0})$
capable to yield the voids observed today \cite{OBAM}
\begin{equation}\label{void4}
\delta (N_{0})\ge 10^{-2}\ \ .
\end{equation}
In Fig.3 we plot again all the
constraints for $N_{2}\ll N_{0}\ll N_{3}$ (upper panel,
negligible post-thin wall and gravitational corrections) and 
$N_{2}=30,\ \ N_{3}=70$ (lower panel).
In both panels, each curve refers to the indicated constraint: the 
solid line represents the condition (\ref{g}), and guarantees that 
the bubbles are expanded to the observed sizes. The short dashed 
line draws the condition (\ref{void4}). Finally, 
the long dashed line represents the strong constraint coming 
from the CMB (\ref{void2},\ref{void3}). The part of solid line
below all the curves satisfies all the constraints, and restricts 
to $N_{0}>55$ the epoch of the phase transition; also, the
gravitational and post-thin wall corrections (important in the
lower panel) allow for a viable parameter set slightly larger
than when they are negligible (upper panel).
For physical parameters on the allowed set, 
the presently large scale observed voids correspond, 
both for size and underdensity, to the TV bubbles nucleated during
inflation, without exceeding the constraints coming from the CMB
isotropy.
 
\section{Conclusion}

We have extended the known results on the stochastic gravitational
background produced by colliding bubbles in two fields models
of first order inflation. In such models, a field performs the
first order phase transition and a second field (that in fourth order
gravity is of gravitational origin) provides the 
inflationary slow rolling. The resulting general phaenomenology is that
the phase transition occurs well before the end of inflation. This has
very non-trivial consequences on the gravitational radiation produced
by colliding bubbles. It is found
that the expansion between the phase transition and the end
of inflation cause an additional redshift for 
the peak frequency of the spectrum with respect to the ordinary 
models of extended inflation; also, the energy carried by the 
gravitational waves is reduced by a fraction equal to the 
density contrast of the (not completely empty) bubbles.
The result is that if first order phase transitions 
occurred during the inflationary era 
(realistically arising from spontaneous 
breaking of very high energy symmetries), 
they should leave next future
detectable peaks in the CGB frequency spectrum.
That detection would be a first experimental verification 
of our very high energy physics theoretical investigations.
In particular, a peak at some frequency explored
by the next Ligo interferometers may be 
explained in terms of a first order phase transition 
occurred typically 10$\div$20 $e-$foldings before the 
reheating at GUT energy scales ($T_{e}\simeq 10^{15}$ GeV).

Moreover, the gravitational radiation produced in recently proposed
models capable of generating the observed large scale voids, has
been examined. In these models, the phase transition occurs
more than 50 $e-$foldings before the end of inflation.
As expected, the typical frequencies of the spectrum
are in the range in which the isotropy of the CMB puts 
the strongest existing constraint on the amplitude of the gravitational
background, and gives a very deep insight into the
physical parameter space. We find a precise new relation among
the physical parameters in the class of the models 
in which the phase transition is fast (namely less than an Hubble 
time); the CMB constraint requires that the phase transition occurs
generally more than 55 $e-$foldings before the end of inflation.
These results localize the parameter set of the effective Lagrangian
for which the present large scale observed voids correspond, 
both in size and underdensity, to the TV bubbles nucleated during the
inflationary era, without perturbing the CMB. The fraction of CMB 
anisotropies and polarization due to phaenomena investigated here 
could be recognized in the high resolution CMB data that will be 
provided by the experiments of the next decade.

\centerline{\bf Figure captions}
\begin{itemize}
\item Fig.1.\ Amplitude-peak frequency relation for the CGB 
generated in a first order phase transition occurred $N_{0}$ 
$e-$foldings before the end of inflation. 
For clearness, we have exploited the two interesting cases 
$\delta_{N_{0}}=$.1 (top) and .01 (bottom) 
and we have evidentiated the dependence of the curves on $N_{0}$
(the grey tonality change correspond to a scansion of 5 in $N_{0}$,
as indicated in the legend box).
Note the shift of the frequencies towards ranges that will be
investigated by the next interferometric detectors.
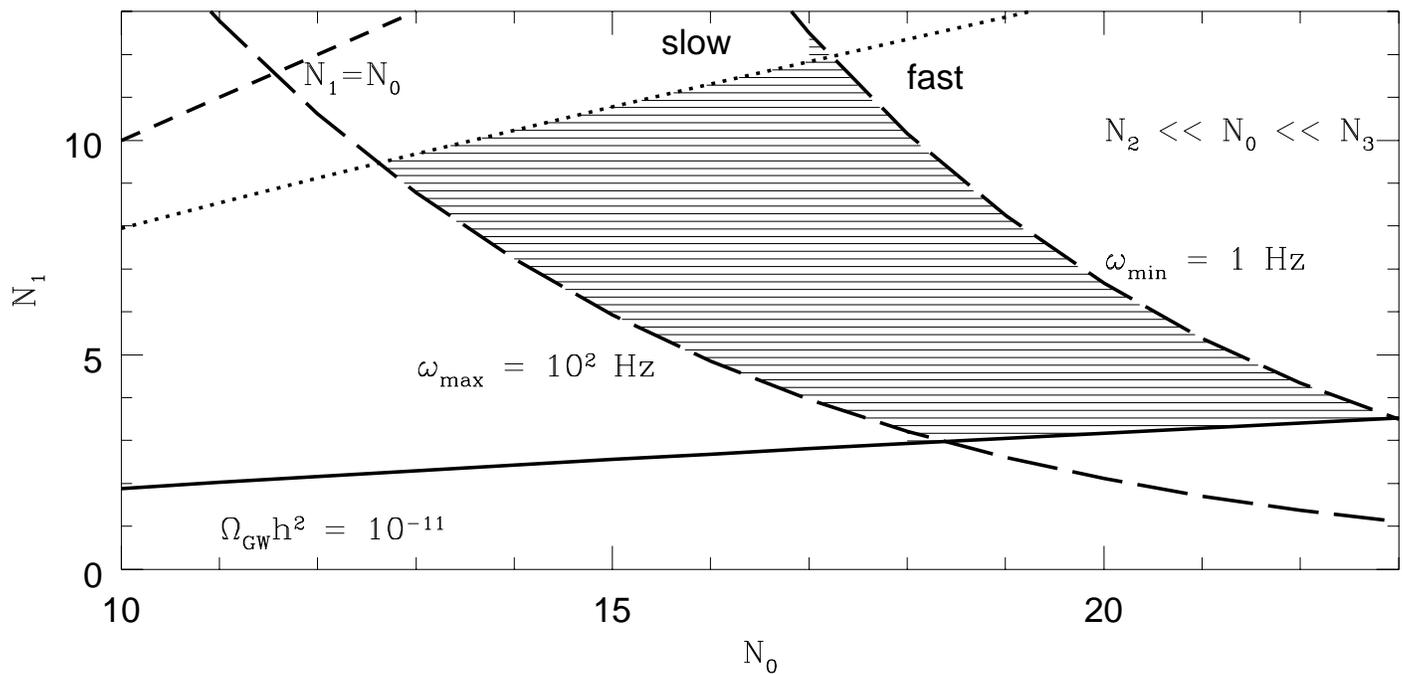
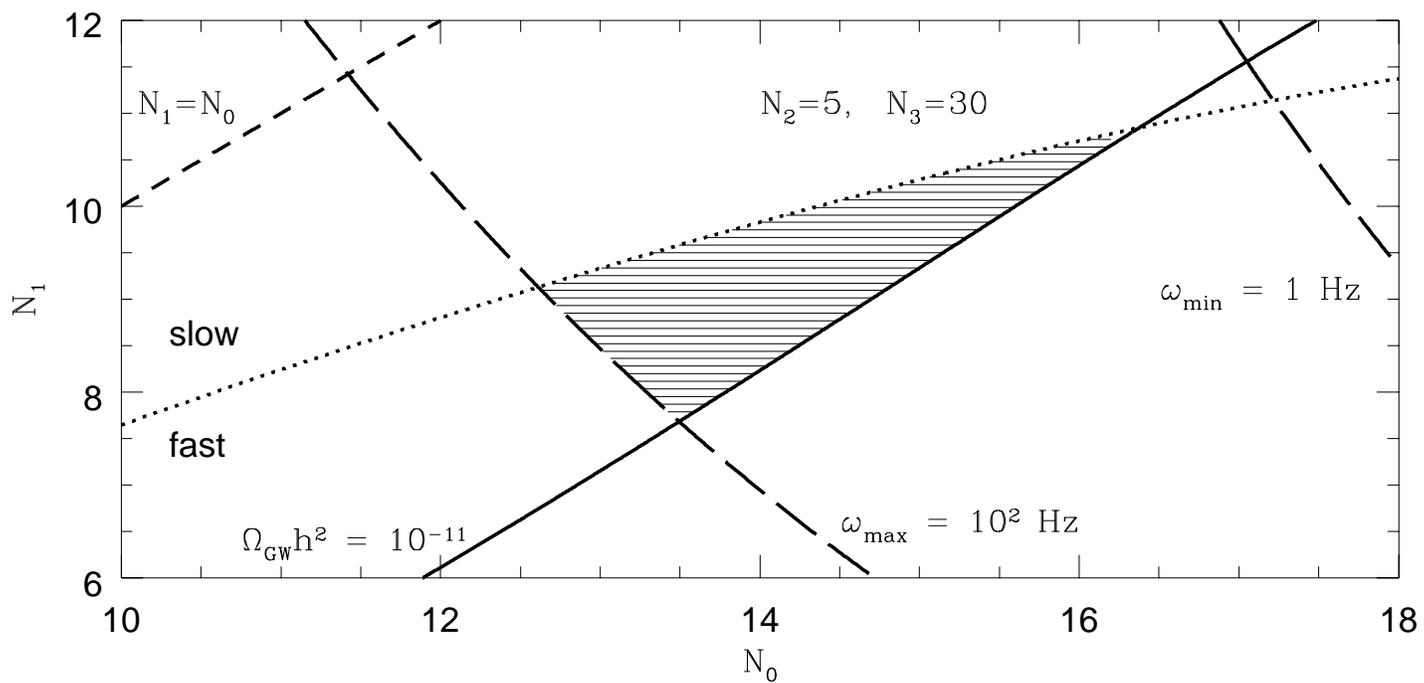
\item Fig.2.\ Advanced Ligo detection region for ($N_{0},N_{1}$), 
in the cases for which the post-thin wall and gravitatinal corrections
are negligible (upper panel) or not (lower panel). 
The dashed areas contain physical 
parameters respecting all the indicated constraints; particularly it 
yelds frequencies among the maximal and minimal limits
(long broken lines) and amplitude above the minimal observable one
(solid lines).
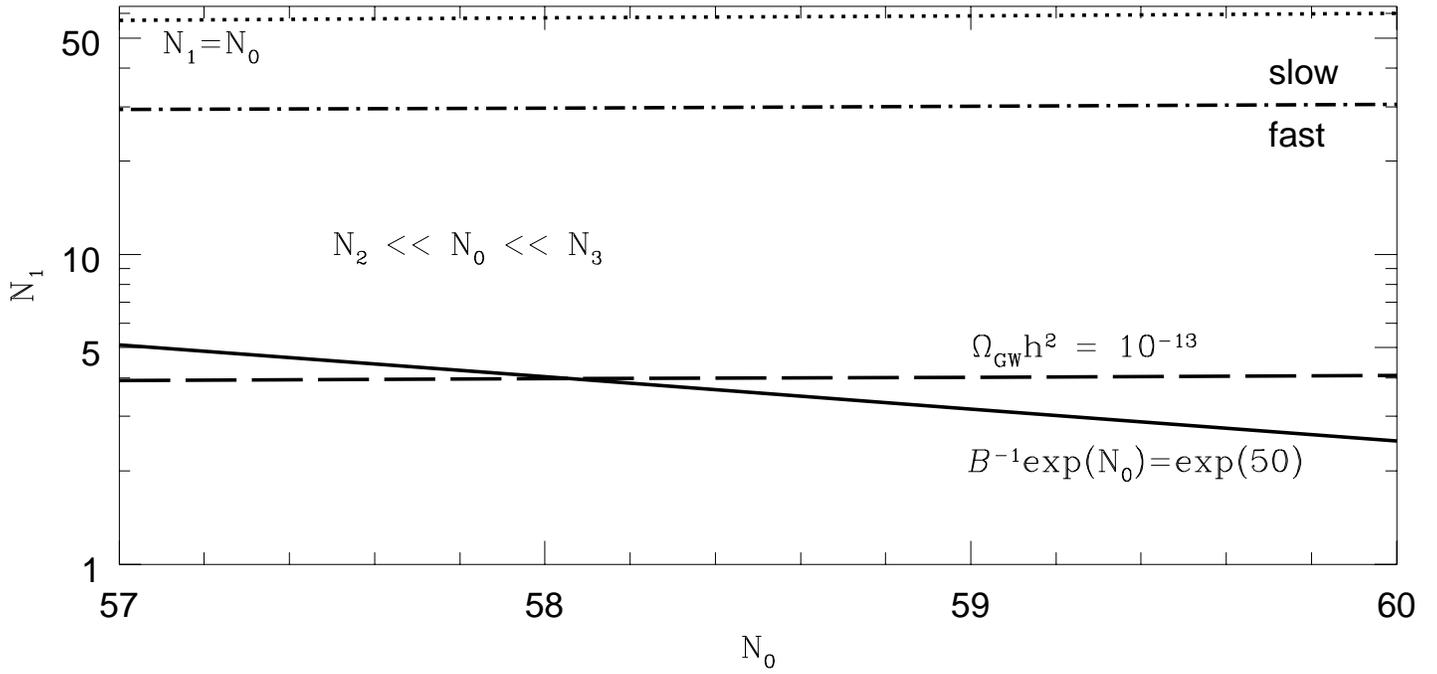
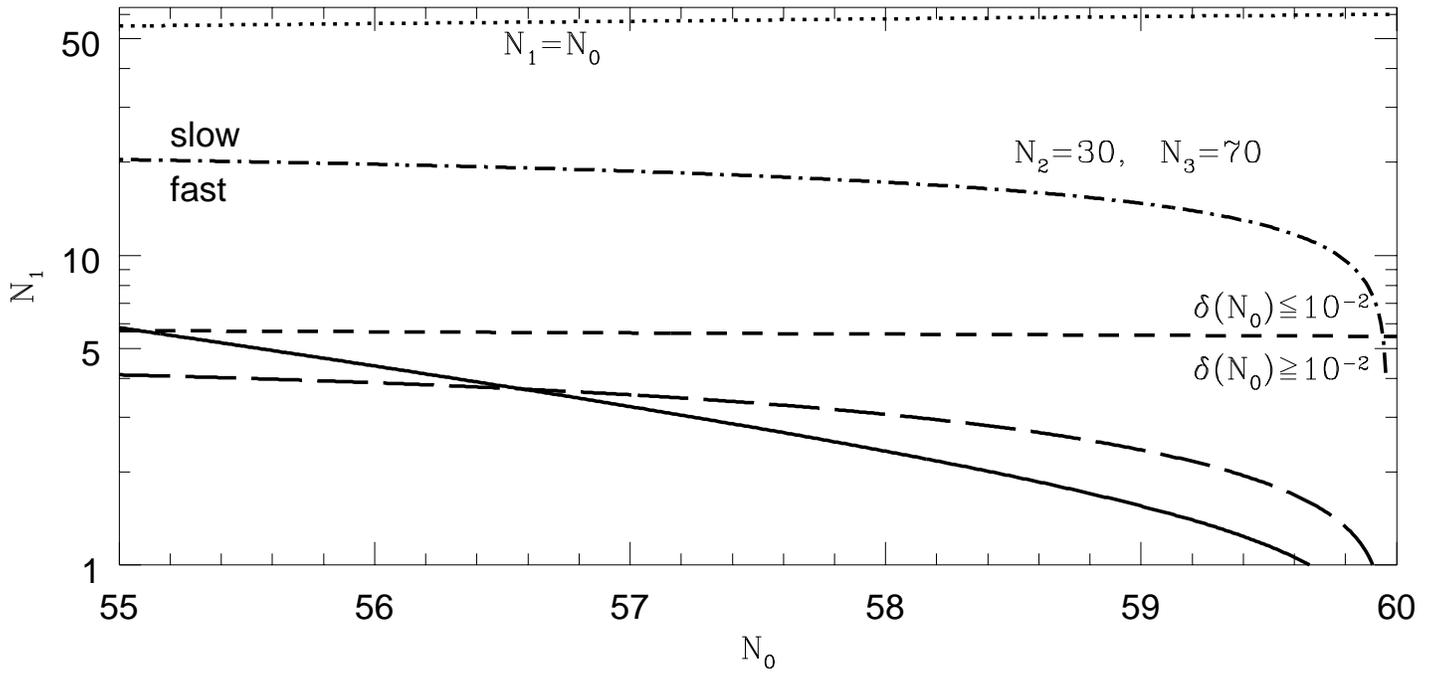
\item Fig.3.\ Insight upon the physical parameters of effective
Lagrangian yielding large scale structure from bubbles; 
the cases for which the post-thin wall and gravitational corrections
are negligible (upper panel) or not (lower panel) are shown.
The segments of the solid line below all the other curves 
sketch the relation among the physical 
parameters for models capable of explaining the large scale observed 
voids as the relics of bubbles nucleated more than 55 $e-$foldings 
before the end of inflation; all the performed constraints are 
respected, and the $N_{1}$ axis is logaritmic to show all them;
note particularly the request of sufficientely deep bubbles (short
broken lines) and the strong constraint coming from the upper limit 
to the level of CMB anisoptropy (long dashed lines).
\end{itemize}

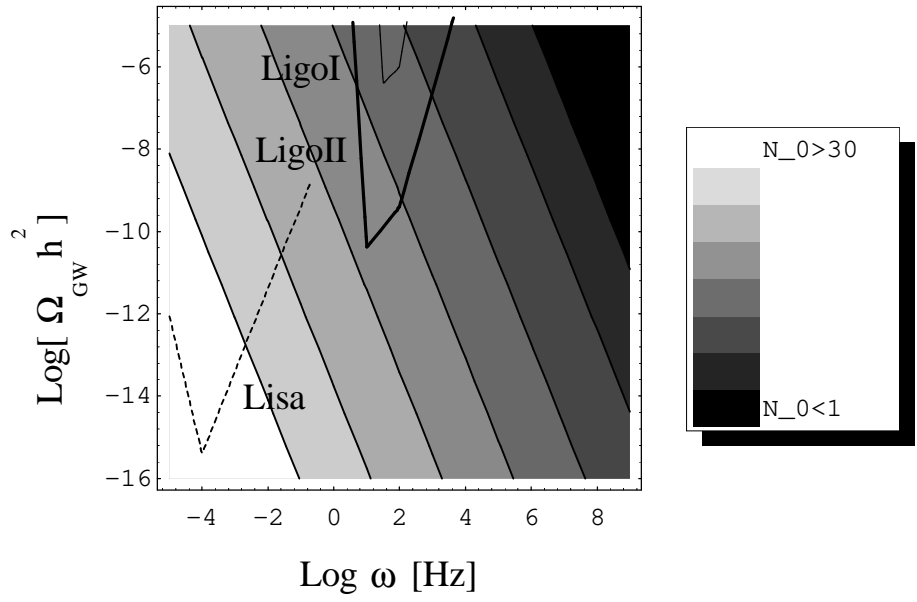
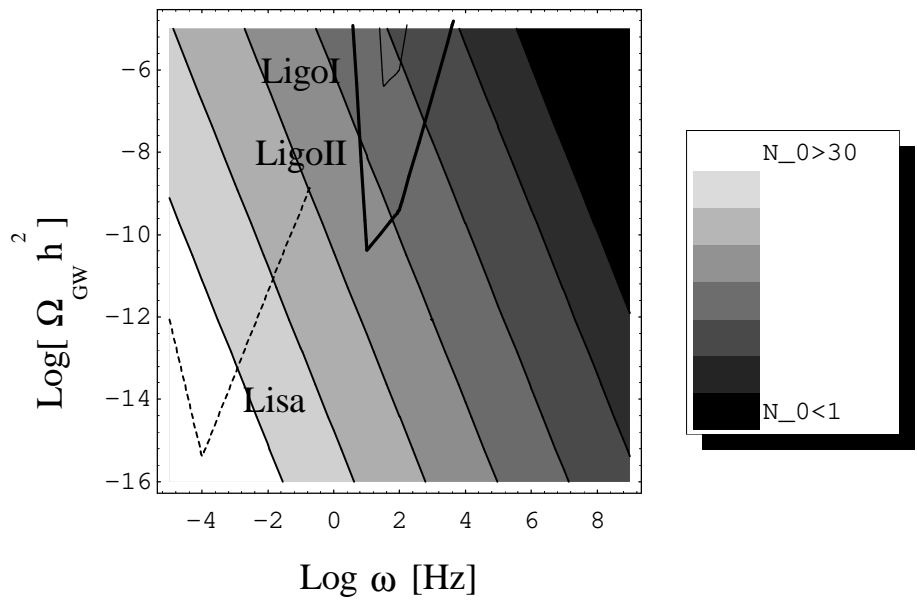
\end{document}